\documentclass[aps, prl, superscriptaddress, reprint, floatfix]{revtex4-1}
\usepackage{graphicx}
\usepackage{amssymb, amsmath}
\usepackage{bm}
\usepackage{hyperref}
\usepackage[all]{hypcap} 

\newcommand{\ev}[1]{{\left<#1\right>}}
\newcommand{\bu}{{\mathbf u}}
\newcommand{\bR}{{\mathbf R}}

\begin{document}

\title{Complex quasi two-dimensional crystalline order embedded in VO$_2$ and
other crystals}

\date{\today}

\author{Timothy Lovorn}
\affiliation{Department of Physics and Astronomy, The University of Alabama,
Tuscaloosa, Alabama 35487, USA}
\affiliation{Department of Physics, The University of Texas at Austin,
Austin, Texas 78712, USA}
\author{Sanjoy K. Sarker}
\affiliation{Department of Physics and Astronomy, The University of Alabama,
Tuscaloosa, Alabama 35487, USA}

\begin{abstract}

Metal oxides such as VO$_2$ undergo structural transitions to low-symmetry
phases characterized by intricate crystalline order, accompanied by rich
electronic behavior. We derive a minimal ionic Hamiltonian based on symmetry
and local energetics which describes structural transitions involving all
four observed phases, in the correct order. An exact analysis shows that
complexity results from the symmetry-induced constraints of the parent phase
which forces ionic displacements to form multiple interpenetrating groups
using low-dimensional pathways and distant neighbors. Displacements within
each group exhibit independent, quasi two-dimensional order, which is
frustrated and fragile. This selective ordering mechanism is not restricted
to VO$_2$: it applies to other oxides which show similar complex order.

\end{abstract}


\maketitle

\emph{Introduction}.---Vanadium dioxide (VO$_2$) undergoes a transition from
a high-symmetry rutile structure to a lower-symmetry monoclinic M1 phase,
with intricate antiferroelectric (AFE) crystalline
order~\cite{morin1959, westman1961, hyland1968, goodenough1971}.
The lowering of symmetry doubles the unit cell, changing the electronic band
structure and converting a metal into a dimerized Mott insulator which shows
unusual metal-insulator coexistence near $T_c$. The insulating phase is
structurally soft, as two other variants appear with doping or application of
strain~\cite{pouget1974, pouget1975, atkin2012, eyert2002_VO2}.
Although such materials are promising for applications (including
low-dissipation logic~\cite{yang2011, nakano2012, zhou2015} and many others),
the lack of a microscopic theory has hindered progress.
VO$_2$ is not alone: similar complex ordering also occurs in many other
crystals~\cite{hiroi2015, eyert2000, eyert2002_NbO2, tanaka2004},
exhibiting diverse electronic and magnetic phases~\cite{imada1998} 
whose properties would certainly reflect the intricacies of the underlying
crystalline order.
To understand exactly how this complexity appears and its effect on finite
temperature properties, one needs a dynamical model to describe the
structural phases.

In this Letter, we focus on VO$_2$ and derive a minimal Hamiltonian in terms
of discrete ionic displacements based on the symmetry of the parent rutile
(R) structure and local energetics.
We find that the model has a broadly applicable selective ordering mechanism
which is responsible for the complexity, which also leads to fragile quasi
low-dimensional order embedded in the 3D system, features that were not
known previously.  
In the R phase of VO$_2$, V ions occupy the sites of a body-centered 
tetragonal lattice surrounded by O octahedra. 
Below a temperature $T_c$ (341 K) a monoclinic (M1) phase emerges in which
V ions are slightly displaced to form dimers, which also twist, creating
zigzag chains along the $c$-axis and giving rise to long-range AFE order.
Another monoclinic phase (M2) and a triclinic one (T) are realized by
modest doping or strain. In the M2 phase only one half of the V chains
dimerizes, while the other half twists, leading to a different
electronic behavior. Rice \emph{et. al.} argued that dimerization in
one chain induces a twist in a neighboring chain and that the M1 phase
is a superposition of two M2 structures~\cite{rice1994}. 
Phenomenological Landau theories based on two order parameters have been
developed~\cite{brews1970, paquet1980} and applied close to
the transition~\cite{tselev2010}.

We show by a detailed mean-field analysis that the microscopic model,
which is exactly mapped into a spin-1 two-component Ising
(Ashkin-Teller~\cite{ashkin1943, ditzian1980}) model,
describes these complex phases. Further (exact) analysis reveals
new features: the complex ordering is quasi two-dimensional, frustrated
and fragile.
These arise due to constraints imposed by the already ordered and densely
packed parent (rutile) phase; the ions make intricate lower-energy
displacements which order separately in several interpenetrating groups
involving planes and distant neighbors. Neighbors belonging to different
groups compete, creating local configurations which tend to frustrate order.
This mechanism is also applicable to other systems showing similar
complex order.

\emph{Ionic model}.---To derive the model, we assume that the potential
energy seen by a V ion at a site $i$ has local minima along the $c$-axis
and apical directions (in the $ab$-plane) located by displacement
vectors $\bu_i$, corresponding to its equilibrium positions in the
rutile ($\bu_i = 0$) and monoclinic phases ($\bu_i \ne 0$).
We construct a Hamiltonian in terms of $\bu_i$, based on the symmetries
of the rutile lattice and local energetics.
This is an effective model since only V ions are considered; $\bu$'s should
be thought of as block variables describing displacements of V ions screened
by the surrounding O ions and electrons.

\begin{figure}
    \begin{center}
    \includegraphics[width=134pt]{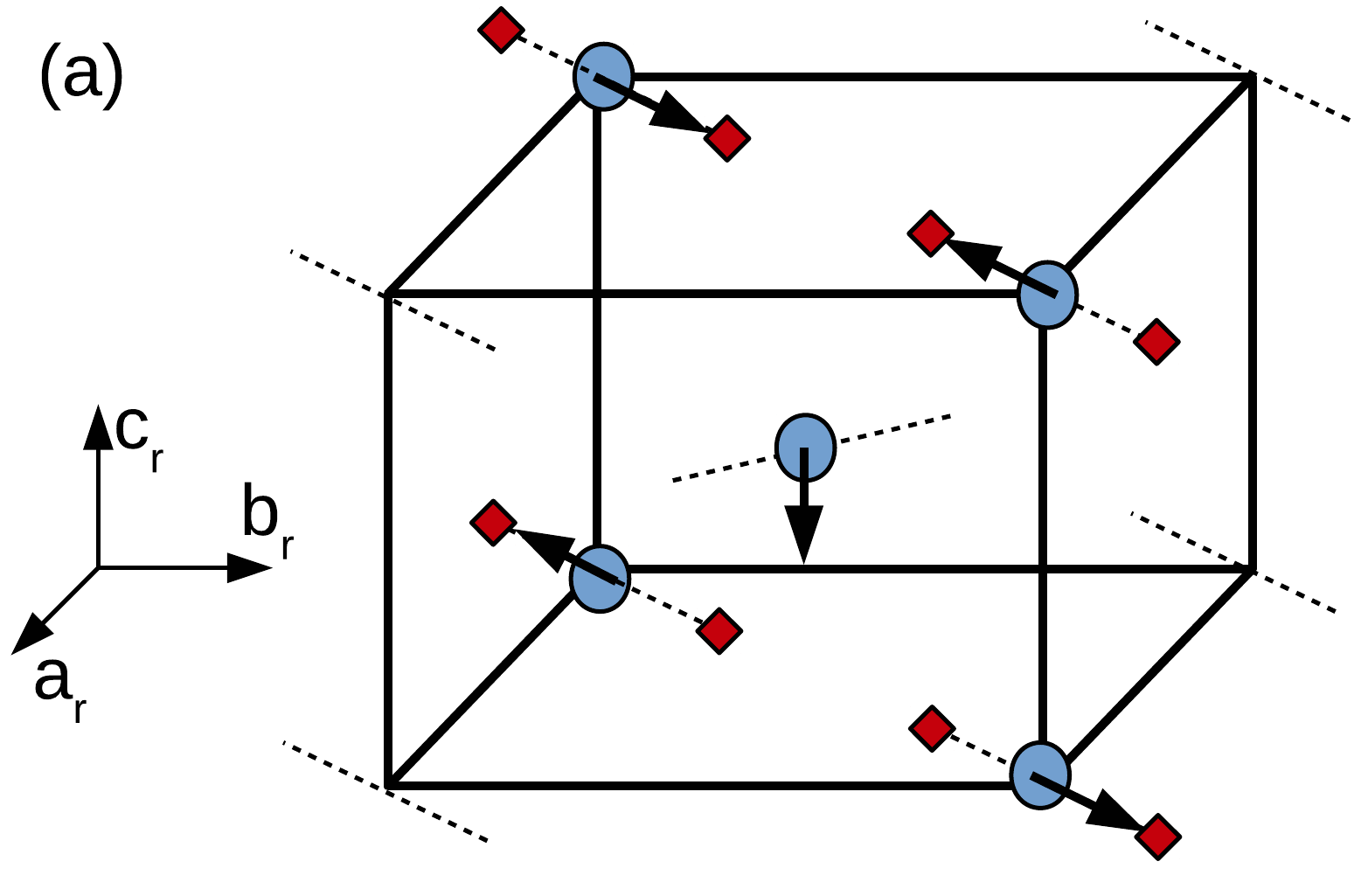}
    \includegraphics[width=106pt]{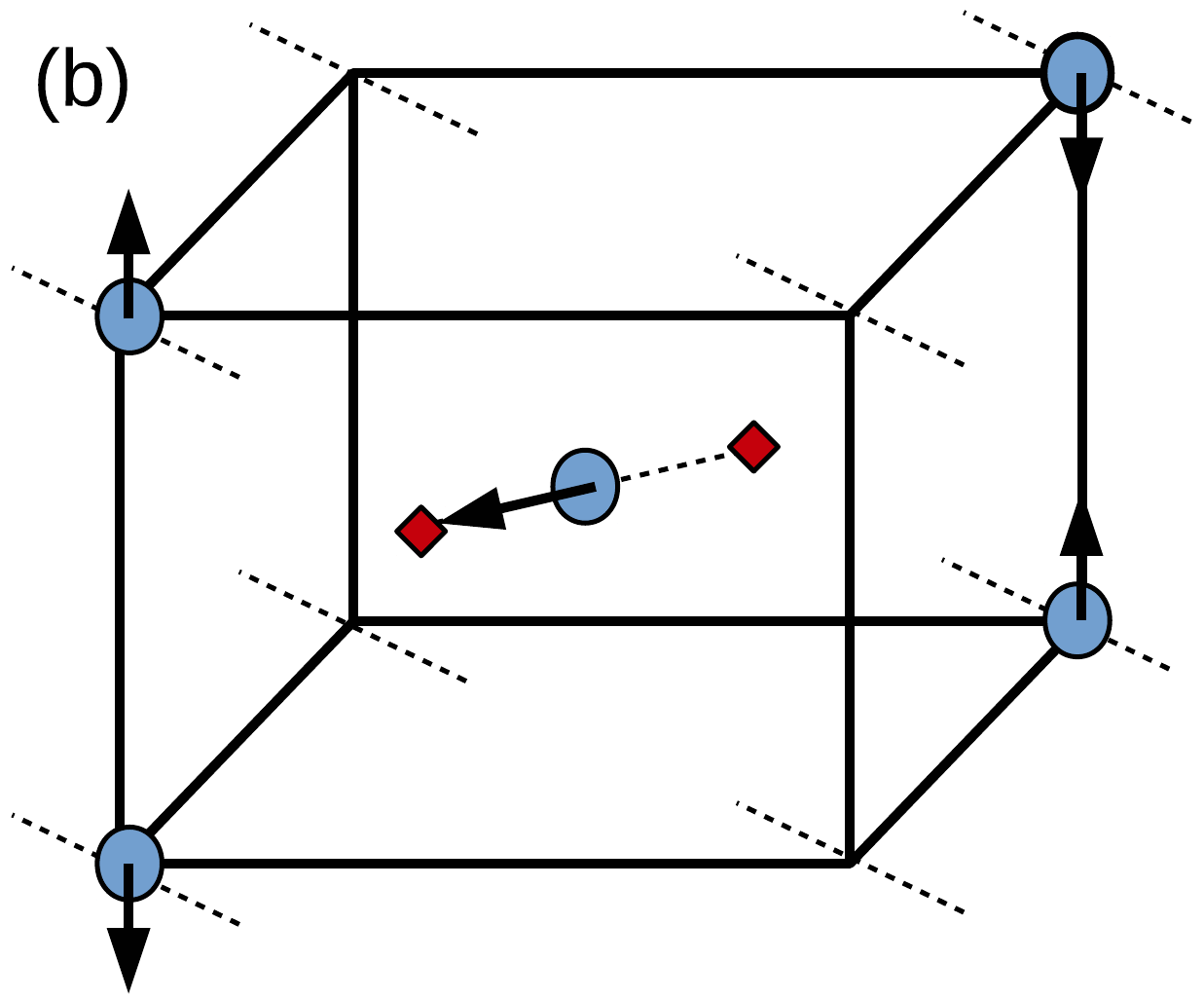}
    \end{center}

    \caption{Interaction between corner and body-center V ions.
    (a) Interaction between V ions (circles) along the $[110]$ direction,
    mediated by corner site apical oxygens (diamonds). Corner V atom arrows
    point towards the low-energy displacement, assuming the indicated
    displacement of the body-center V atom. The rutile lattice
    vectors $\mathbf{a}_r$, $\mathbf{b}_r$, $\mathbf{c}_r$ are shown for
    reference.
    (b) Corresponding interaction between V ions along the $[1\bar{1}0]$
    direction, mediated by body-center site apical oxygens.}
    \label{fig:modes}
\end{figure}

The rutile phase has an \emph{orientational order} which plays the central
role in determining the Hamiltonian. The oxygen octahedra have two different
orientations connected by a screw operation.
Four O ions form a V-centered rectangle, with one side along the $c$-axis
and the other on the $ab$-planes along one of the diagonals.
All body center V ions have rectangles oriented one way, along $[110]$,
as shown in Fig.~\ref{fig:modes}(a).
All corner V ions have them in a perpendicular direction,
along $[1\bar{1}0]$, as shown in Fig.~\ref{fig:modes}(b).
We denote the sets of planes in which these rectangles lie as A and B,
respectively.

Consider a body-center V ion.
The two O$^{2-}$ ions below it lie along the diagonal of the $ab$-plane
joining two corner V$^{4+}$ ions.
The effective V-O attraction is clearly the strongest in this
configuration; the minima for corner V ions, if they exist, will be along
this line.
Therefore, $\bu$ has only one component, $u_{ab}$, in the $ab$-plane.
There are no O ions, and hence no minima, along the other diagonal.
When the two O ions move away from each other, they pull the two corner V
ions along the $ab$-diagonal toward each other and push the two body-center
V ions above and below away from each other vertically (by electrostatics).
Hence, $\bu$ has another component $u_c$ along the $c$-axis.
Thus, a twist (by $u_{ab}$) causes an out-of-phase dimerization (by $u_c$)
via an interaction mediated by the O-pair (and O-V bonding electrons),
as observed. 

Each component has 3 values by mirror symmetry, and thus can be represented
by a (pseudo) spin-1 Ising variable:
$u_{ic} = u_z S_i$, $u_{iab} = u_x \sigma_i$, where $S$ and $\sigma$ take
values $(0, \pm 1)$; $u_x$ and $u_z$ are positive (we use the convention
$u_{iab} > 0$ when moving in the $+\hat{x}$ (${\mathbf a}_r$)
direction from site $i$).
Thus we have an important result: structural transitions are described by
a spin-1 two-component Ising (Ashkin-Teller) model.
The rich critical properties exhibited by the spin-1/2 Ashkin-Teller model
have been extensively studied.
With 9 on-site states the spin-1 model can describe much more complex
behavior~\cite{badehdah1999, santos2016}, but it has not been studied widely.
Its realization in VO$_2$ and other crystals reflects the potential
complexity of the latter's behavior.

Using the mapping we can derive the Hamiltonian $H = H_0 + H_{int}$, where
\begin{equation}
    H_0 = \sum_i \left[b_z S_i^2 + b_x \sigma_i^2 
    + b_{xz} S_i^2 \sigma_i^2\right]
\end{equation}
is the most general form of the on-site term since $S^3 = S$ and
$\sigma^3 = \sigma$.
The sum is over all sites $i$.
There are no odd terms by symmetry.
$H_{int}$ describes intersite interactions, characterized by energies
$J_{ij}$ which we assume decrease rapidly with distance.
The $b$'s and $J$'s arise from effective ionic interactions.
Consider a tetragonal cell with the body-center spins labeled
($S_b$, $\sigma_b$), and the corners labeled by
coordinates $(x, y, z = 0,1)$.
The dominant interactions are between the dimerizing spin ($S_b$)
and twisting corner spins ($\sigma$), and vice versa.
Keeping these, we have
\begin{equation}
\begin{split}
    H_{int} =& -J_b \sum \left[S_b (\sigma_{000} - \sigma_{110}
    - \sigma_{001} + \sigma_{111})\right.\\
        & \left.+ \sigma_b (S_{010} - S_{100} - S_{011} + S_{101})\right]
\end{split}
\label{eq:H_int}
\end{equation}
with $J_b > 0$.
The sum is over cells. 
We show that the Hamiltonian $H_0 + H_{int}$ describes the phases of VO$_2$.
The lowest interaction energy corresponds to  
$\sigma_{000} = -\sigma_{110} = -\sigma_{001} = \sigma_{111} = S_b$,
with $S_b = 1$ or $-1$ in the first bracket, 
and similarly in the second with $S$ and $\sigma$ interchanged, which gives 
$E_0 = b_x + b_z + b_{xz} - 4J_b$ for the energy per site. 
This is the M1 phase with spins oppositely directed (antiferroelectric
ordering) along the $c$- and the $ab$-diagonals.
It is stable at $T = 0$ if $E_0 < 0$.

To determine the phase diagram as a function of temperature, we use a
mean-field approximation. Since corners
and body centers are nonequivalent, there are four order parameters:
the pairs $\ev{S_i}$, $\ev{\sigma_i}$ for the corner and the body center
sites, where $\ev{\dots}$ denotes thermal averaging. 
Suppose we label the spins by the sites $\bR$ of a tetragonal lattice 
and a two-point basis $(\bR,p)$, 
with $p = c$ for the origin (corner) and $p = b$ for the body center.
Since the interaction $H_{int}$ is antiferroelectric, the order parameters
can be chosen as
\begin{equation}
    (\ev{S_{\bR p}},\ev{\sigma_{\bR p}}) = (m_{Sp},m_{\sigma p})
    e^{i \mathbf{Q}_a \cdot \bR}
\end{equation}
where $\mathbf{Q}_{a} = (\pi/a,0,\pi/c)$, and $a$, $c$ are the lattice
spacings. We have solved the four coupled    
mean-field equations for the four order parameters
numerically~\footnote{The
software developed to solve the mean-field equations is available at
\url{https://github.com/tflovorn/vo2mft}
and \href{https://doi.org/10.5281/zenodo.195417}{doi:10.5281/zenodo.195417}.}.
Going back 
to the original notation, we see that a body-center $S_i$ is coupled to the
nearest corner $\sigma_j$'s within the same plane (A). It follows
that $m_{sb}$ is nonzero only if $m_{\sigma c}$ is nonzero, and vice versa.
Thus, the group comprising body-center $S$ spins and corner $\sigma$ spins
in plane A is effectively characterized by a single order parameter;
we represent it by the average $m_A = (m_{sb} + m_{\sigma c})/2$.
Similarly, body-center $\sigma$ spins interact only with corner $S$ spins
in the B plane, forming a second group characterized by the order
parameter $m_B = (m_{sc} + m_{\sigma b})/2$.
Although the two groups interact through the on-site quartic term, the
two order parameters can exist independently since if one is zero, the
other need not be. Each describes a twist and a connected dimerization.

The M1 phase, with $m_A = m_B$, is stable at low $T$.
If $b_{xz} = 0$, the two groups are decoupled; each is described by a
(one-component) spin-1 Ising model.
This well-known (Blume-Emery-Griffiths~\cite{blume1971}) model has been
used to describe He$_3$-He$_4$ mixtures and other systems. 
For $b_x = b_z$, the Blume-Capel phase diagram~\cite{blume1966, capel1966}
is reproduced.
A line of second-order transitions from M1 to R phase up to
$b_x/4J_b \approx 0.46$ is followed by a line of first-order transitions
up to $b_x/4J_b = 0.5$, which is important since the observed transition
is first order in VO$_2$.

\begin{figure}
    \begin{center}
    \includegraphics[width=240pt]{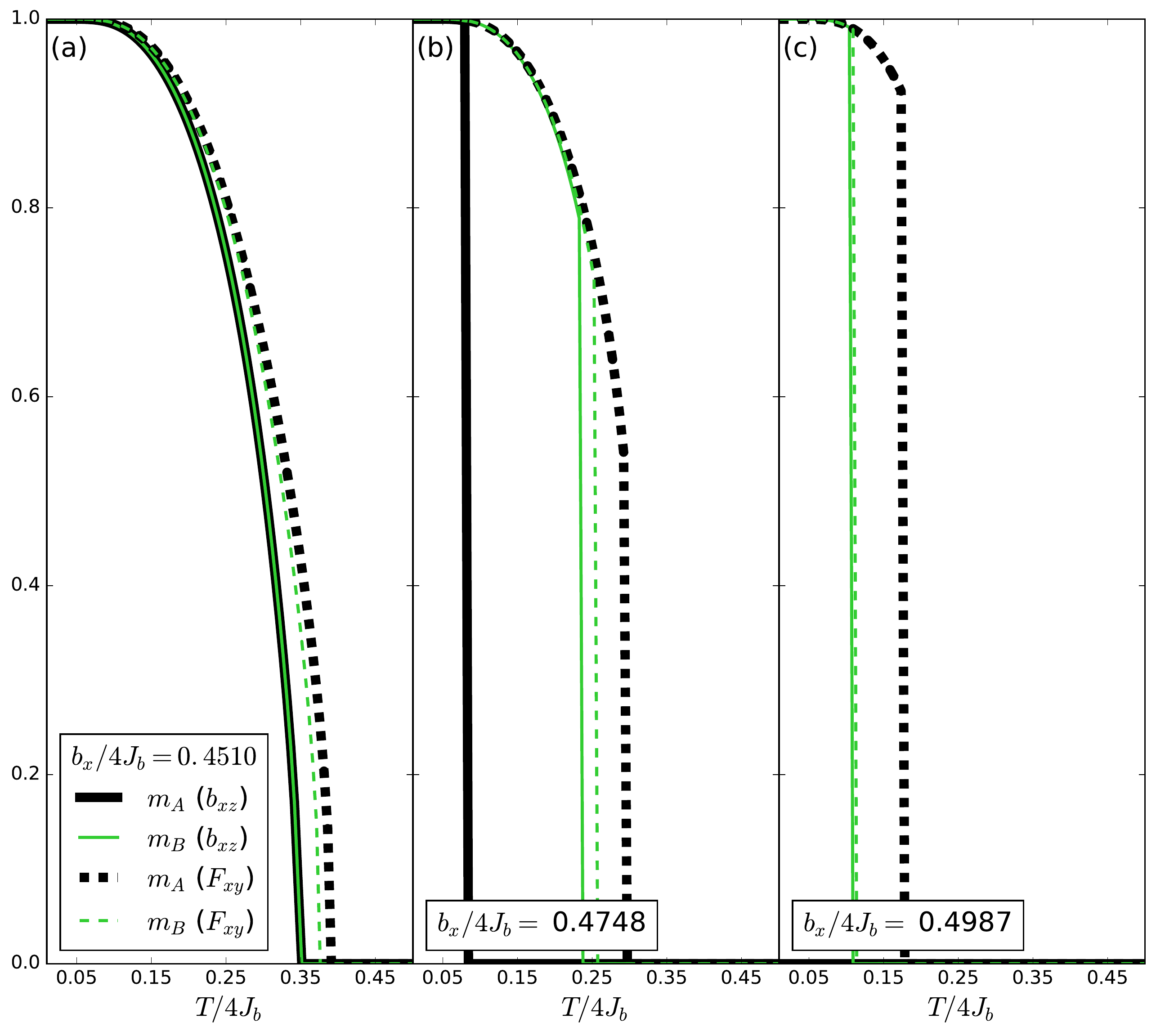}
    \end{center}

    \caption{Mean-field order parameter and phases.
    (a) Order parameters $m_A$ and $m_B$ including the quartic
    interaction $b_{xz}$ (solid lines) or the screw axis symmetry
    breaking $F_{xy}$ (dashed lines) are shown in the regime of small
    on-site quadratic term $b_x/J_b$ (taking $b_z = b_x$). The M1 phase
    occurs at low temperature for the $b_{xz}$ case, while the T phase
    develops in the $F_{xy}$ case; a second-order transition to the R phase
    occurs with increasing temperature.
    (b) Same as in (a) but for larger on-site quadratic term.
    First-order transitions are present between M1, M2, and R phases
    ($b_{xz}$ case) and T, M2, and R phases ($F_{xy}$ case).
    (c) Same as in (b) but with further increased $b_x/J_b$.
    In the $b_{xz}$ case, the M2 phase is present at zero temperature.}
    \label{fig:mft_order_parameters}
\end{figure}

The quartic term ($b_{xz} > 0$) represents a competition between
on-site twisting and dimerization.
The system compromises by stabilizing the M2 phase, with
$m_A \neq 0$ and $m_B = 0$ (or vice versa), between the M1 and the R phases.
This is shown by the solid lines in Fig.~\ref{fig:mft_order_parameters}
for $b_{xz}/J_b = 0.1$.
There are three regimes: one with second-order transitions between
M1 and R phases, one with first-order transitions between M1 and M2
and between M2 and R, and one in which the M1 phase is absent.  

The M2 phase can also be stabilized by applying strain along
the $[110]$ direction~\cite{pouget1975}.
This breaks the screw axis symmetry, making $b_x$ different for corner
and body sites, which we parameterize by
$b_{x,corner} = (1 - F_{xy})b_x$ with $F_{xy} \ll 1$.
In this case a triclinic (T) phase with $m_B < m_A \neq 0$ also appears
between the M1 and M2 phases, as shown by the dashed lines
in Fig.~\ref{fig:mft_order_parameters} where $F_{xy} = 0.03$.
The transitions between T and M2 phases are second order for small $b_x/J_b$,
and elsewhere they are first order.
All of these results are qualitatively consistent with the observed phase
diagram, which provides an important experimental validation of the
correctness of the model. 
Landau free energies used in phenomenological
theories~\cite{brews1970, paquet1980, tselev2010} can be obtained by
expanding the mean-field free energy in powers of the order parameters.

\emph{3D ordering and frustration}.---The mean-field approximation
(and Landau theories) may hide an important property of the 
Hamiltonian $H_0 + H_{int}$, which can be seen from
Eq.~\ref{eq:H_int} and Fig.~\ref{fig:modes}.
Any twisting spin $\sigma_b$ interacts with only half (4) of the corner $S$
spins which are all in \emph{one} A plane, and the dimerizing spin $S_b$
interacts with the 4 corner $\sigma$ spins which are all in \emph{one}
B plane.
In each case, the coupling with the other 4 corner spins is quartic
(thus, not symmetry breaking) since the displacements are perpendicular.
Therefore, the spins are divided into two groups.
The body-center $S$- and corner $\sigma$-spins live on one set of parallel
planes (A) with no quadratic (symmetry breaking) coupling between the planes.
The other group, with $(S, \sigma)$ interchanged, live on the perpendicular
set of planes (B) and have the same property. 
Therefore, the Hamiltonian is invariant under the 2D global symmetry
transformation: $S_i \rightarrow -S_i$, $\sigma_i \rightarrow -\sigma_i$
independently on each plane.
Hence, each plane orders independently with the order parameter
either $m$ or $-m$.
The ground state is $2^{2L}$-fold degenerate, where $L$ is the number
of planes in each direction.
These exact results remain valid in the presence of intra- and inter-plane
quartic (e.g. $S_i^2 \sigma_j^2$) interactions.

\begin{figure}
    \begin{center}
    \includegraphics[width=240pt]{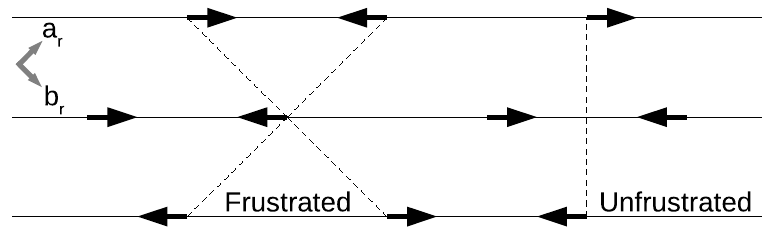}
    \end{center}

    \caption{Interaction between planes leading to fragile 3D order.
    Arrows represent the $ab$-plane ($\sigma$) degree of freedom on corner
    sites. 2D AFE order is established by the strong interaction between
    these degrees of freedom along the $[110]$ axis, which is shown as a
    solid line and mediated by the body-center $c$-axis ($S$) degree of
    freedom. The interactions along this axis form the A planes as shown 
    in Fig.~\ref{fig:modes}(a). Interactions between adjacent A planes,
    indicated by the dashed lines on the left side, are frustrated: the
    2D order within the A planes prevents this type of interaction from
    minimizing its energy. Weaker interaction between next-nearest-neighbor
    A planes, indicated by the dashed line on the right side, is required
    to establish 3D order. A similar situation exists for the B planes which
    form along the $[1\bar{1}0]$ axis.}
    \label{fig:interaction_planes}
\end{figure}

To see how 3D ordering may develop we need to consider other quadratic
interactions which have been neglected so far, e.g., of the form
$-\sum J_{ij} S_i S_j$, where $S_i$ and $S_j$ are in different planes. 
The dominant ones are again between body-centers and corners (see
Fig.~\ref{fig:modes}). 
By symmetry, all $J_{ij}$ are the same.
The interaction has the same form as Eq.~\ref{eq:H_int},
but with all $S$ spins and all with positive signs.
Hence, its thermal average value in the ordered phases vanishes,
irrespectively of the sign of $J_{ij}$.
This is also true for $\sigma$ spins, and for second-neighbor
interactions (in the $ab$-planes), since quite generally, for an $S_i$
in one plane, the $S_j$'s in an adjacent plane add up to zero in pairs
for an AFE ordered phase. Hence these interactions do not lead to 3D
ordering.
On the contrary, they prefer ferroelectric order \emph{within}
the planes and therefore are frustrated. 
This interaction is shown by the dashed lines (forming triangles) on the left
side of Fig.~\ref{fig:interaction_planes}.

However, interactions between spins in alternate parallel planes lead
to 3D ordering, as they do not compete with the in-plane order since there
is no pairwise cancellation. The strongest of these couples nearest spins on
alternate planes in the perpendicular direction, and is of the form:
\begin{equation}
    H_{3D} = -J_{3D,S} \sum S_i S_j - J_{3D,\sigma} \sum \sigma_i \sigma_j
\end{equation}
This interaction is shown by the dashed line on the right side of
Fig.~\ref{fig:interaction_planes}. It leads to 3D ordering within 
the set of A and/or B planes,
which is ferroelectric if $J_{3D}$ is positive, and antiferroelectric
if it is negative. Since only alternate planes are coupled, each set
consists of two interpenetrating subsets which are ordered independently, 
so that the ground state is now $2^4 = 16$-fold degenerate,
corresponding to choices of $\pm m_A$ and $\pm m_B$. These results are exact.
Since $J_{3D}$ actually corresponds to third-neighbor interaction,
$J_{3D}/J_b$ is small and 3D ordering weak, and consequently each subset
shows quasi 2D order.

This degeneracy is robust against two-body perturbations, as the arguments above and
the (isosceles) triangular construction in Fig.~\ref{fig:interaction_planes}
can be generalized to any
pair of planes with arbitrary range of interaction $J_{ij}$.
Then, the two spins forming the base of the triangle (in the second plane)
are aligned if the planes are in the same subset because the base length equals
an even number of lattice spacings.
It is odd when the two planes are in different subsets, so that the two spins
are anti-aligned, i.e, there is no ordering between the subsets in the
dimerized phase.
Ordering can be established (and degeneracy reduced) by making $J_{ij}$ strong
enough; but that will destroy the dimers, causing a transition to another
(ferroelectric) phase.

Since the 16-fold degenerate order is protected against two-body perturbations,
one may consider selective multi-body interactions which, though weaker, are
allowed by symmetry.
An example is $K (S_1 S_2)_A (S_3 S_4)_B$, where the four spins belong to
four different planes, say, two body-center spins of A type and two corner
spins of B type, in the most compact arrangement.
For $K > 0$ this term has a lower energy when the ordering wave vectors
for A and B planes are either both ${\mathbf Q}_a = (\pi/a, 0, \pi/c)$
or both ${\mathbf Q}_b = (0, \pi/a, \pi/c)$, thereby lowering the
degeneracy to 8.
For $K < 0$ the corresponding choice is ${\mathbf Q}_a$ for one type of
plane and ${\mathbf Q_b}$ for the other.

In short, the 3D order is not only weak, it is characterized by several
energy scales with different degrees of degeneracy which will appear as
different crossover scales for $T > 0$.
The various 3D ordering interactions have to be included to study these
effects.
The frustrating interactions between spins belonging to opposite
(interpenetrating) subsets are generally stronger than the more
distant-neighbor ordering interactions $J_{3D}$ and the four-spin
interactions $K$.
While they do not change the degeneracy structure, their effects,
which are not included in the mean-field (or Landau) theories,
are important for $T > 0$ since they provide an extra source of
entropy and also weaken the in-plane order.
Together with the complex degeneracy including the hidden $2^{2L}$-fold 2D
degeneracy, they make the ordered state quite fragile and prone to breaking
up into domains.

Our main results are not restricted to VO$_2$. The model itself applies to
any rutiles, particularly to oxides of the form MO$_2$
(where M is a transition metal).
At least five of these (M = Mo, W, Tc, $\alpha$-Re, Nb) have similar
low-energy paired (dimerized)
structures~\cite{hiroi2015, eyert2000, eyert2002_NbO2};
all are predicted to have similar selective frustrated ordering. 
The materials which do not undergo the transition are also described
by the model.
Nor is the mechanism restricted to rutiles. 
For example, Ti$_2$O$_3$ (more generally, M$_2$O$_3$~\cite{imada1998})
with a primary corundum structure shows a transition to a monoclinic
phase with similar complex paired structure~\cite{hiroi2015, tanaka2004}.
In this case, the pseudospin model may be different but can be derived
using our template.
Thus we have found a generic mechanism by which an important class of ionic
crystals develop complex and fragile secondary crystalline order using
selective low-dimensional pathways. 
The nature of the diverse electronic and magnetic states exhibited by these
materials is expected to be rather different from those of usual crystals.
The model provides a foundation for studying these by introducing
residual electron-ion interactions~\cite{tanaka2004, paquet1980}.

\begin{acknowledgments}
This research was supported by the National Science Foundation (DMR-1508680).
Work in Austin was supported by the Department of Energy, Office of Basic
Energy Sciences under contract DE-FG02-ER45118 and by the Welch Foundation
under grant TBF1473.
\end{acknowledgments}

%

\end{document}